\documentclass{article}

\usepackage[margin=1in]{geometry}

\usepackage{graphicx}
\usepackage{xspace}
\usepackage{amsmath}
\usepackage{amssymb}
\usepackage{amsthm}
\usepackage{graphicx}
\usepackage[square,sort&compress,numbers]{natbib}
\usepackage{ upgreek }
\usepackage{float}
\usepackage{url}
\usepackage{algorithm}
\usepackage[noend]{algpseudocode}
\usepackage{subcaption}
\usepackage[colorinlistoftodos]{todonotes}

\graphicspath{{./}{./Figures/}}

\theoremstyle{definition}
\newtheorem{definition}{Definition}

\theoremstyle{example}
\newtheorem{example}{Example}

\theoremstyle{observation}
\newtheorem{observation}{Observation}

\def\Per{\mathrm{Per}}
\def\Fix{\mathrm{Fix}}

\def\Acyc{\ensuremath{\mathrm{Acyc}}}

\title{Attractor Stability in Finite Asynchronous Biological System Models}
\author{
Henning S.~Mortveit$^1$ \and
Ryan Pederson$^2$
}

\date{
	$^1$Engineering Systems and Environment and Network Systems Science \& Advanced Computing, University of Virginia, Charlottesville, VA, USA \\ \texttt{Henning.Mortveit@virginia.edu}\\%
	$^2$Department of Physics and Astronomy, University of California, Irvine, CA, USA \\ \texttt{pedersor@uci.edu}\\[2ex]%
}

\begin{document}
\maketitle

\begin{abstract}
We present mathematical techniques for exhaustive studies of long-term dynamics of asynchronous biological system models. Specifically, we extend the notion of $\kappa$-equivalence developed for graph dynamical systems to support systematic analysis of all possible attractor configurations that can be generated when varying the asynchronous update order (Macauley and Mortveit (2009)). We extend earlier work by Veliz-Cuba and Stigler (2011), Goles et al. (2014), and others by comparing long-term dynamics up to \emph{topological conjugation}: rather than comparing the exact states and their transitions on attractors, we only compare the attractor structures. In general, obtaining this information is computationally intractable. Here, we adapt and apply combinatorial theory for dynamical systems from Macauley and Mortveit (2008, 2009, 2011, 2012, 2014) to develop computational methods that greatly reduce this computational cost. We give a detailed algorithm and 
apply it to~($i$) the~\emph{lac} operon model for~\emph{Escherichia coli} proposed by Veliz-Cuba and Stigler (2011), and~($ii$) the regulatory network involved in the control of the cell cycle and cell differentiation in the~\emph{Caenorhabditis elegans} vulva precursor cells proposed by Weinstein et al. (2015). In both cases, we uncover all possible limit cycle structures for these networks under sequential updates. Specifically, for the \emph{lac} operon model, rather than examining all~$10! > 3.6 \cdot 10^6$ sequential update orders, we demonstrate that it is sufficient to consider~$344$ representative update orders, and, more notably, that these~$344$ representatives give rise to \emph{$4$ distinct attractor structures}. A similar analysis performed for the \emph{C.~elegans} model demonstrates that it has precisely~$125$ distinct attractor structures. We conclude with observations on the variety and distribution of the models' attractor structures and use the results to discuss their robustness. \\


\end{abstract}

\section{Introduction}
\label{sec:intro}
Here we will assume that biological models are represented as maps of the form 
\begin{equation} 
  F = (f_1,\dots,f_n) \colon K^n \longrightarrow K^n \;,
\label{eq:F}
\end{equation}
where~$K$ is some suitable finite set like~$\{0,1\}$, see for
example~\cite{goles2013neural}. We refer to maps of the form~\eqref{eq:F} as discrete~\emph{graph dynamical systems} (GDSs).
Associated to~$F$ we have its~\emph{dependency graph}~$G$ with vertex set~$V(G) = \{1, 2,\dots,n\}$ and edges all~$\{i, j\}$ for which the function~$f_j$ depends non-trivially on the vertex state~$x_i$, see~\cite{adiga2015network}. The map~\eqref{eq:F} may be assembled through parallel update of states (the synchronous case), or through asynchronous update methods such as sequential or block sequential~\cite{goles2013deconstruction}. With this system representation, questions about an underlying biological system can be turned into precise questions about the dynamical system defined by the map~$F$ in~\eqref{eq:F} using the structure of~$G$ and the properties of the maps~$f_i$. 

In this paper we focus on the attractors of the map in Equation~\eqref{eq:F}; they capture the possible long-term behaviors of a system model. In the context of a biological system, the attractors will often represent some physical observable such as the stages of a cell cycle. Naturally, it is desirable that the long-term dynamics of a model reflects the associated experimental observations.  
  
In this paper, we demonstrate our mathematical techniques for exploring the possible attractor structures using the following biological network models: the regulatory network involved in the control of the cell cycle and cell differentiation in the~\emph{Caenorhabditis elegans} vulva precursor cells proposed by~\cite{weinstein2015model}, and the~\emph{lac} operon in~\emph{Escherichia coli} proposed by~\cite{veliz2011boolean}. These models are examples of GDSs; more details surrounding these networks are presented in Section~\ref{sec:models}. The authors in both cases evaluated their models using a synchronous (or parallel) update method and found their models to exhibit dynamical behavior that is consistent with published experimental results. 

For a synchronous model, the assumption is that all genes make a transition at the same time. While there is seldom enough kinetic information to discern a precise order of state transitions~\cite{garg2008synchronous}, the assumption of synchronized transitions may not be biologically realistic. By relying solely on analysis derived using a synchronous update scheme, one runs the risk of missing important insight on the structural stability of the model, see for example~\cite{Demongeot:10}. To assess how one's model behaves under perturbations can offer valuable insight into its structural stability which in turn can help justify the extent to which one's conclusions are generic.
For this reason, it is advantageous to examine how a biological model behaves under asynchronous update schemes such as sequential update orders. In general, obtaining this information across all sequential update orders is computationally intractable as here the number of updates grows as~$n!$, where~$n$ is the number of vertices in the model. Although existing programs, such as BoolNet ~\cite{BoolNet} and GenSim ~\cite{GenSim}, can be used to compute the attractors from a sequential update, it would be extremely costly to perform a brute force analysis over all $n!$ sequential updates. 

In this paper we present an algorithm for analyzing the possible dynamics resulting from all the possible sequential update orders. Other update schemes are discussed in Section~\ref{sec:conc}. Our analysis provides biological modelers immediate feedback on how their model behaves under different asynchronous updating orders and the diversity in attractor structures. Clearly, a large degree of attractor structure sensitivity with respect to the update scheme raises a cautionary flag regarding the validity and soundness of the modeling effort. 

The concept of network robustness is presented in~\cite{kitano2004biological} as a uniquitously observed and necessary property of biological systems. Naturally, one may also consider systems capable of evolution, in which robustness is present up to some point. The purpose of this paper is not to consider robustness versus evolution of the system being modeled, but instead to \emph{offer the modeler a systematic way to examine model robustness with respect to the update scheme.}  
For this, we say that a dynamical property is robust when it is not affected by small perturbations. In a real system, these perturbations can model the inherent noise present in nature, whereas in a model, perturbations refer to changes in the state variables or to changes in the specification of the model itself~\cite{aracena2009robustness}. In this paper we will consider perturbations of the update order. Unlike perturbations to the system state, these modifications represent changes to the model itself and therefore may cause variations in the dynamics of the system. Here we only consider variations across different sequential (permutation) update order. For more information on robustness of update orders see~\cite{Macauley:10,Macauley:12a,Macauley:14,aracena2009robustness}.

Several biological network models have previously been studied in the context of update order robustness such as the~\emph{lac} operon model~\cite{montalva14attraction}, mammalian cell cycle network model~\cite{ruz2014dynamical}, and yeast cell cycle network~\cite{goles2013deconstruction}. In these papers, the authors study update schedule robustness by analyzing dynamics properties such as long-term behavior (fixed points and limit cycles) and attractor basin sizes across all fixed deterministic updating schemes. 
We extend and complement this work by presenting computational algorithms for identifying all possible attractor structures that can be obtained for sequential update orders. At a high level, this algorithm is based on the body of work in~\cite{macauley2009cycle,Macauley:14,Macauley:12a,Macauley:10} with key facts outlined below. It assumes that we have a fixed list of vertex functions~$f = \{f_i\}_{i=1}^n$ capturing the local evolution with associated dependency graph~$G$. All definitions are carefully presented in Section~\ref{sec:back}.

\medskip\noindent\rule[0mm]{1.5mm}{1.5mm} For an update sequence $\pi = (\pi_1, \ldots, \pi_n)$ of the vertices of~$G$, we obtain a map $F_\pi$ of the form~\eqref{eq:F} by applying the maps $f_i$ in the sequence specified by $\pi$, see Equation~\eqref{eq:sds}.

\smallskip\noindent\rule[0mm]{1.5mm}{1.5mm} From~\cite{reidys1998acyclic} we have the equality $F_\pi = F_{\pi'}$ whenever the acyclic orientations~$O_\pi$ and~$O_{\pi'}$ of~$G$ are equal. All possible maps of the form $F_\pi$ can therefore be obtained from $\Acyc(G)$, the set of all acyclic orientations of~$G$: for each acyclic orientation $O$, pick a sequence $\pi$ compatible with $O$ (a linear extension of $O$). 

\smallskip\noindent\rule[0mm]{1.5mm}{1.5mm} From the main result in~\cite{macauley2009cycle} we know that the maps~$F_\pi$ and~$F_{\pi'}$ have the same attractor structure whenever the corresponding acyclic orientations $O_\pi$ and $O_{\pi'}$ are \emph{$\kappa$-equivalent}, that is, are related by a sequence of source-to-sink operations. We remark that this is an equivalence relation.

\smallskip\noindent\rule[0mm]{1.5mm}{1.5mm} Using results from~\cite{Macauley:08b}, we show how to construct a complete set of update sequence representatives for $\kappa$-equivalence. From this set of representatives, we can generate all possible attractor structures of maps of the form~$F_\pi$. 

\smallskip\noindent\rule[0mm]{1.5mm}{1.5mm} We also show how to compute~$\kappa(G)$, the number of $\kappa$-classes of~$G$. The quantity~$\kappa(G)$ is an upper bound for the number of possible attractor structures that can be generated, and it is typically orders of magnitude smaller than $n!$, the total number of update orders $\pi$. For example, in the \emph{lac} operon network we have~$\kappa(G) = 344$ which is four orders of magnitude less than~$10!$.

\medskip

The algorithm that we present is a concise, computational recipe for constructing a complete set of update sequence representatives for attractor equivalence. This is a prime example of a \emph{structure-to-function} result: the theory and algorithm are based entirely on the combinatorial structures of the dependency graph~$G$.

\medskip

\noindent\textbf{Significance.} In our approach, we have coarsened the analysis to consider the variety and distribution of cycle structures across different sequential updates, rather than supplying full details about the complete dynamics of the model. This analysis may point to clear inconsistencies and instabilities resulting from the model choice. For example, a biological system may be known to only have steady states (fixed points). If a proposed model of a biological system exhibits limit cycles as attractors, that may be an indication that the model is not valid or has shortcomings. Our method helps reduce the computational burden related to conducting this and other analyses for asynchronous models. 

\medskip

\noindent\textbf{Paper outline.} We present basic terminology and theory in Section~\ref{sec:back} followed by a short overview of regulatory network models for~\emph{C. elegans} and~\emph{lac} operon in Section~\ref{sec:models}. Methods and algorithms are presented in Section~\ref{sec:method} with computational illustrations presented in Section~\ref{sec:results}. Here we demonstrate our method by identifying all limit cycle structures in the~\emph{C. elegans} and~\emph{lac} operon network models in the case of sequential update orders. To the best of our knowledge, results pertaining to the former model have not been published, whereas partial results for the latter model have appeared in~\cite{montalva14attraction}. Concluding remarks appear in Section~\ref{sec:conc}.

\section{Background, Terminology, and Notation}
\label{sec:back}

The theory covered in this section is a careful assembly of existing results that are needed for deriving our algorithms~\cite{reidys1998acyclic,Macauley:14,Macauley:11c,macauley2009cycle}. A sound understanding of these results is essential to using the algorithms and techniques effectively. 
We consider discrete dynamical systems as in Equation~\eqref{eq:F} where each map~$f_i$ is of the form~$f_i \colon K^n \longrightarrow K$. In general, each~$f_i$ will only depend non-trivially on some subset of the \emph{vertex states}~$x_1$ through $x_n$, a fact captured by the dependency graph~$G$ defined earlier. Each vertex~$i$ of~$G$ has assigned a~\emph{vertex state}~$x_i \in{K}$ and a~\emph{vertex function} of the form~$f_i \colon K^{d(i)+1} \longrightarrow K$ taking as arguments the states of vertices in the closed 1-neighborhood of~$i$ in~$G$. Here~$d(i)$ is the degree of vertex~$i$. We write~$n[i]$ for the ordered sequence of vertices from the~$1$-neighborhood of~$i$ (note that~$i$ included), and~$x[i]$ for the corresponding sequence of vertex states, see~\cite{adiga2015network}. The \emph{system state} is the $n$-tuple consisting of all the vertex states, and is denoted by~$x = (x_1,\dots,x_n) \in{K^n}$. 

\begin{figure}[!t]
\centerline{      \fbox{\includegraphics[scale=0.333]{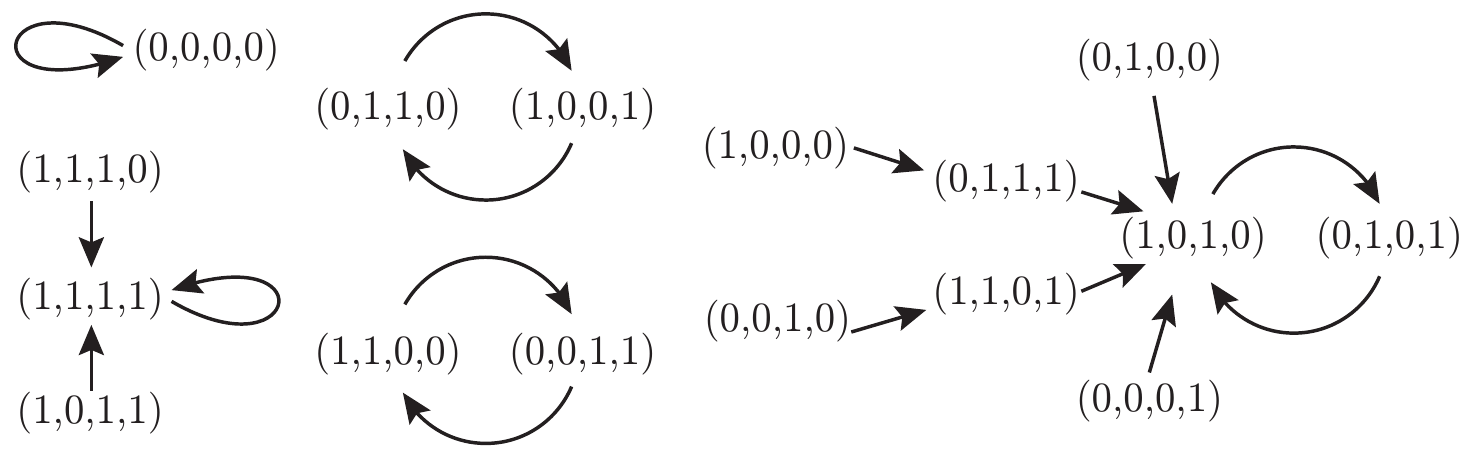}}\quad
 \fbox{\includegraphics[scale=.666]{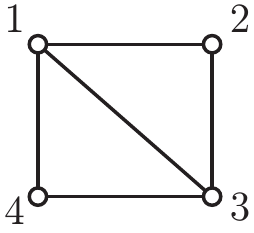}}\quad
  \fbox{\includegraphics[scale=.333]{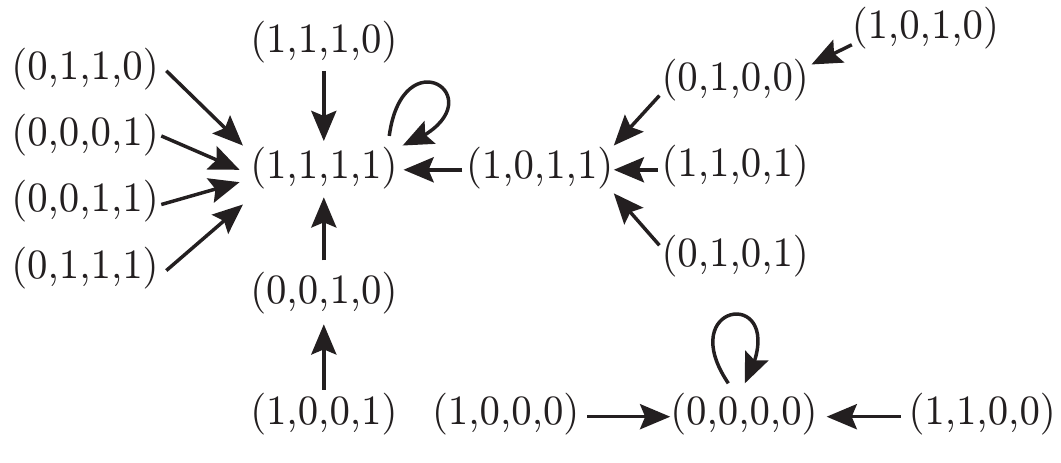}}
}
\caption{The graph~$G$ of Example~\ref{ex:psEX} is shown in the middle. The phase space of the map~$F$ for the parallel update is shown on the left while the phase space of the map~$F_\pi$ obtained under a sequential update order~$\pi = (1,2,3,4)$ is shown on the right. Clearly the update scheme matters. This figure is used with permission from~\cite{kuhlman2014attractor}.}
\label{fig:psExample}
\end{figure}

If we apply the functions $(f_i)_i$ \emph{synchronously (in parallel)} we obtain the graph dynamical system map~$F$ as in Equation~\eqref{eq:F} given by
\begin{equation}
  F(x_1,\dots,x_n) = \bigl(f_1(x[1]),\dots,f_n(x[n])\bigr) \, .
\end{equation}
However, in this paper we will consider permutation update sequences, that is, we will apply the functions~$f_i$ in the order given by a permutation~$\pi = (\pi_1, \ldots, \pi_n)$. For this we first introduce the notion of~$G$-\emph{local functions}~\cite{kuhlman2011bifurcations}. Here the~$G$-local function~$F_i \colon K^n \longrightarrow K^n$ is given by
\begin{equation}
  F_i(x_1,\dots,x_n) = (x_1,x_2,\dots, f_i(x[i]),\dots,x_n) \, .
\end{equation}
Using the permutation~$\pi \in{S_G}$ (the set of all permutations of~$V(G)$) as an update sequence, the corresponding~\emph{sequential dynamical system} (SDS) map~$F_{\pi} \colon K^n \longrightarrow K^n$ as in Equation~\eqref{eq:F} is given by
\begin{equation}
\label{eq:sds}
  F_{\pi} = F_{\pi(n)} \circ F_{\pi(n-1)} \circ \dots \circ F_{\pi(1)} \, . 
\end{equation}
The~\emph{phase space} of the dynamical system map~$F$ in~\eqref{eq:F} is the directed graph~$\Upgamma(F)$ with vertex set~$K^n$ and directed edges all pairs~$(x,F(x))$. A state on a cycle in~$\Upgamma(F)$ is called a~\emph{periodic point} and a state on a cycle of length one is called a~\emph{fixed point}~\cite{kuhlman2011bifurcations}. The sets of all such points are denoted by~$\Per(F)$ and~$\Fix(F)$, respectively. The collection of all the cycles in the phase space represents all possible long-term dynamics of the system model. 
\begin{definition}
Let~$F$ be as in Equation~\eqref{eq:F}. The~\emph{cycle structure} of~$F$ is the unlabeled subgraph of~$\Upgamma(F)$ induced by~$\Per(F)$. 
\end{definition}
We represent the cycle structure of a phase space using \emph{multiset} notation so that, e.g.~$\{1(2), 2(3)\}$ denotes a phase space with two fixed points and three~$2$-cycles as in Example~\ref{ex:example} and shown on the left side in Figure~\ref{fig:psExample}. 
For illustrations, we will use Boolean~\emph{bi-threshold} vertex functions~\cite{kuhlman2014attractor} $\mathbf{t}_{i,k^\uparrow,k^\downarrow,m} \colon \{0,1\}^m \longrightarrow \{0,1\}$ defined by
\begin{equation}
\mathbf{t}_{i,k^\uparrow,k^\downarrow,m}(x_1,\dots,x_m) =
\begin{cases}
1, &\text{if } x_i = 0 \text{ and } \sum_{j=1}^{n} x_j \geq k^\uparrow,\\
0, &\text{if } x_i = 1 \text{ and } \sum_{j=1}^{n} x_j < k^\downarrow,\\
x_i, &\text{otherwise},\\
\end{cases}
\end{equation}
where the integer parameters~$k^\uparrow$ and~$k^\downarrow$ are the~\emph{up-threshold} and~\emph{down-threshold}, respectively. 
\begin{example}
\label{ex:example}\label{ex:psEX}
To illustrate the above concepts, let~$G$ be the simple graph shown in the middle of Figure~\ref{fig:psExample} and let each vertex function be given by a bi-threshold function with thresholds~$(k^\uparrow,k^\downarrow) = (1, 3)$. The (synchronous) map~$F$ has the phase space shown on the left in Figure~\ref{fig:psExample}. Using~$\pi = (1, 2, 3, 4)$ as the update sequence, we obtain the map~$F_\pi$ with phase space as shown on the right in Figure~\ref{fig:psExample}. This example is taken from~\cite{kuhlman2014attractor}.
\end{example} 
The dependency graph~$G$ is generally directed and will contain loops (edges of the form~$(v,v)$.) Let~$G_c$ be the graph obtained from~$G$ by~($i$) removing all self-loops,~($ii$) replacing each directed edge by an undirected edge, and~($iii$) collapsing all parallel edges to a single edge. For all the theory in this paper, we obtain exactly the same results and conclusions whether we use~$G$ or~$G_c$. 
To see this, assume that~$G$ is directed and possibly also contains loops. The questions of determining whether~$F_\pi$ and~$F_{\pi'}$ over~$G$ are functionally- or cycle-equivalent are precisely captured by the graph~$G_c$ as before: consider the two compositions $F_a \circ F_b$
and $F_b \circ F_a$ in the case of functional equivalence. The conditions for when these compositions are equal are independent of edges in~$G$ being directed or not, or if there are loops. We simply state that this argument transfers to and holds for~$\kappa$-equivalence, but refer the reader to the proof for cycle equivalence of~$F_\pi$ under cyclic shifts of the update sequence given in~\cite{macauley2009cycle}. The upshot of all this is that for a given biological model, \emph{we may simply consider the $G_c$ versions of their respective networks}. When referring to these graphs later in this paper, we will not distinguish between~$G$ and~$G_c$; it will be clear from the context. We call~$G_c$ the \emph{combinatorial graph} induced by~$G$.
 


\subsection{Functional Equivalence} 
The first step to address is: for update sequences $\pi$ and $\pi'$, when is $F_\pi = F_{\pi'}$? 
To answer this, ~\cite{reidys1998acyclic} introduced the equivalence relation~$\sim_\alpha$ on~$S_G$ by~$\pi \sim_\alpha \pi'$ if~$\pi$ can be transformed into~$\pi'$ by a sequence of adjacent transpositions of vertices~$v$ and~$v'$ where~$\{v,v'\}$ is not an edge in~$G$. It is clear that~$\pi \sim_\alpha \pi'$ implies the functional equality~$F_\pi = F_{\pi'}$, see~\cite{Macauley:11c}. The equivalence classes are called~$\alpha$-classes, the~$\alpha$-class containing~$\pi$ is denoted by~$[\pi]_G$, and the collection of all~$\alpha$-classes are written~$S_G/\!\!\sim_\alpha$. 

Although it may be natural to address functional equivalence through~$\sim_\alpha$, it is computationally more efficient to use the \emph{acyclic orientations} of~$G$. 
An orientations of~$G$ is an assignment of direction to each edge~$e\in E(G)$. This can be represented by a map~$O_G \colon E(G) \longrightarrow V(G) \times V(G)$; the orientation~$O_G$ is \emph{acyclic} if this assignment contains no cycles. The set of \emph{acyclic orientations} of~$G$ is denoted~$\Acyc(G)$, and we set~$\alpha(G) = |\Acyc(G)|$. 
Here is the connection: a permutation~$\pi\in S_G$ defines an acyclic orientation~$O(\pi)$ by~$O(\{i,j\}) = (i,j)$ if~$i$ precedes~$j$ in~$\pi$ and~$O(\{i,j\}) = (j,i)$ otherwise, see~\cite{Macauley:11c} and Figure~\ref{fig:orientationExample}, left. Moreover, from~\cite{reidys1998acyclic} we have a bijection between the $\alpha$-classes of~$G$ and the acyclic orientations of~$G$ given by
\begin{equation}
\label{eq:phi_map}
\phi_G \colon S_G/{\sim_\alpha} \longrightarrow \Acyc(G)\; , \qquad \phi_G([\pi]_G)= O(\pi) \;.
\end{equation}
The number of equivalences classes under~$\sim_\alpha$ therefore equals~$\alpha(G)$, and it follows that the measure~$\alpha(G)$ is an upper bound for the number of functionally distinct maps~$F_\pi$ that can be generated by varying the update order.

\smallskip\noindent\textbf{Fact~1. Testing for equality of sequential GDS maps.}~We can test if~$\pi$ and~$\pi'$ are in the same~$\alpha$-class, which would imply~$F_\pi = F_{\pi'}$, by computing and comparing~$O(\pi)$ and~$O(\pi')$. The complexity of this test is~$\mathcal{O}(m)$ where~$m = |E(G)|$. 

\smallskip\noindent\textbf{Fact~2. Bounding the number of distinct sequential GDS maps.}~We can compute this bound~$\alpha(G)$ through the recursion relation 
\begin{equation}
\alpha(G) = \alpha(G/e) + \alpha(G\setminus e) \;, \qquad e \in E(G) \;.
\end{equation}
Here~$G/e$ is the graph constructed from~$G$ by contracting the edge~$e$, and~$G\setminus e$ is the graph obtained from~$G$ by deleting the edge~$e$. 

\begin{example} For the graph~$G$ in Example~\ref{ex:example} we have~$\alpha(G) = 18$. To see this, take~$e = \{1,3\}$ so that~$\alpha(G/e) = 4$ and~$\alpha(G\setminus e) = 14$. If we take~$\pi = (1,2,3,4)$ we get the acyclic orientation shown on the left in Figure~\ref{fig:orientationExample}.
\begin{figure}[h]
 \centering
  \begin{subfigure}{0.2\linewidth}      \fbox{\includegraphics[scale=.74]{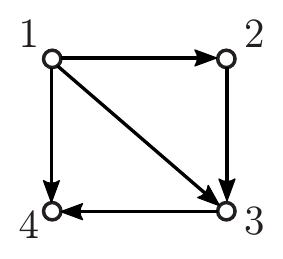}}
  \end{subfigure}
 \qquad
 \begin{subfigure}{0.2\linewidth}
        \fbox{\includegraphics[scale=.74]{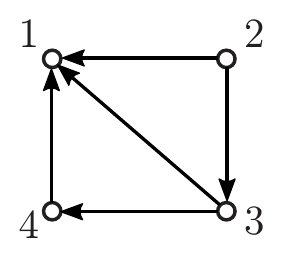}}
    \end{subfigure}
\caption{For the graph~$G$ in Example~\ref{ex:psEX} with~$\pi = (1,2,3,4)$, the orientation~$O(\pi)$ is shown on the left (vertex~$1$ is the unique source). On the right, the acyclic orientation of the shifted order~$\sigma(\pi) = (2, 3, 4, 1)$ is shown. Notice these two orientations are click-related as we have converted vertex~$1$ from a source to a sink.}
\label{fig:orientationExample}
\end{figure}
\end{example}

\subsection{Cycle Equivalence}
Two dynamical systems of the form~\eqref{eq:F} are \emph{cycle equivalent} if they have the same cycle structure. 
Here we specialize this to the case where we have update sequences~$\pi$ and~$\pi'$ and corresponding maps~$F_\pi$ and~$F_{\pi'}$. First, define the cyclic shift~$\sigma(\pi)$ by~$\sigma(\pi) = (\pi_2,\pi_3,\dots,\pi_n,\pi_1)$. At the level of acyclic orientations, we see that mapping~$\pi$ to $\sigma(\pi)$ corresponds precisely to converting~$\pi_1$ from a source in~$O(\pi)$ to a sink in~$O(\sigma(\pi))$, see Figure~\ref{fig:orientationExample}.
Following~\cite{Macauley:11c}, we call the conversion of a source vertex to a sink vertex in~$O \in{\Acyc(G)}$ a~\emph{source-to-sink operation}, or a~\emph{click}. Two acyclic orientations~${O,O'} \!\in\!\Acyc(G)$ where~$O$ can be transformed into~$O'$ by a sequence of clicks are said to be click-related. The transitive and reflexive closure of this relation is denoted by $\sim_\kappa$; it is an equivalence relation, see~\cite{Macauley:11c}. Its equivalence classes are called~$\kappa$-classes, are denoted by~$[O]_G$, and the set of all $\kappa$-classes is~${\Acyc(G)/\!\!\sim_\kappa}$. 

We introduced $\kappa$-equivalence in~\cite{macauley2009cycle} because~$O(\pi) \sim_\kappa O(\pi')$ implies that the maps~$F_{\pi}$ and~$F_{\pi'}$ are cycle equivalent. 
How do we efficiently determine if~$O(\pi) \sim_\kappa O(\pi')$? For this, let~$C$ be a directed simple cycle in~$G$ of length~$k$, that is,~$C = (v_0, v_1,\dots,v_{k-1},v_0)$ where~$\{v_{i-1},v_i\} \in E(G)$ for~$i = 1,\ldots, k$. 
Define the (scalar)~\emph{Coleman $\nu$-function}~\cite{coleman1989killing} 
\begin{equation}
\nu_{C} \colon \Acyc(G) \longrightarrow \mathbb{Z} \; ,
\end{equation}
by~$\nu_{C}(O) = n^+ - n^-$ where $n^+$ is the number of edges in~$C$ oriented along~$O$ and~$n^-$ is the number of edges in~$C$ oriented opposite of $O$, see~\cite{Macauley:11c}. 
We need to extend $\nu_C$ to cover all the cycles of~$G$, and for this 
we pick a \emph{cycle basis}~$\mathcal{C} = (C_1,\dots,C_m)$ and define $\nu_\mathcal{C} \colon \Acyc(G) \longrightarrow \mathbb{Z}^m$ by
\begin{equation}
\nu_\mathcal{C} = (\nu_{C_1}, \ldots, \nu_{C_m}) \;, 
\end{equation}
see~\cite{Macauley:11c}, where it is also shown that~$\nu_{\mathcal{C}}$ is a \emph{complete invariant} for $\kappa$-equivalence. In other words,~$\nu_{\mathcal{C}}(O) = \nu_{\mathcal{C}}(O')$ if and only if~$O \sim_\kappa O'$. We summarize this and more in the following facts that form the basis for the algorithm presented in Section~\ref{sec:method}.

\smallskip\noindent\textbf{Fact 3. Testing for cycle equivalence of sequential GDS maps.} To determine if~$\pi$ and~$\pi'$ are~$\kappa$-equivalent, and therefore if~$F_\pi$ and~$F_{\pi'}$ are cycle-equivalent, ($i$) pick a cycle basis~$\mathcal{C}$ of~$G$;~($ii$) form~$O(\pi)$ and~$O(\pi')$, and~$(iii)$ compare~$\nu_\mathcal{C}(O(\pi))$ and~$\nu_\mathcal{C}(O(\pi'))$.  

\smallskip\noindent\textbf{Fact 4. Exhaustive exploration of all possible attractor structures for sequential GDS maps.} Let~$\Acyc_v(G)$ denote the set of all acyclic orientations of~$G$ where~$v$ is the unique source, and let $f=(f_i)_i$ be fixed vertex functions. To exhaustively explore all possible cycle structures for maps of the form~$F_\pi$, we use the fact that for any~$v \in V(G)$ the set~$\Acyc_v(G)$ is a complete set of~$\kappa$-equivalence class representatives, see~\cite{macauley2009cycle}. The corresponding set of update sequence representatives is obtained by selecting precisely one linear extension for each element of~$\Acyc_v(G)$.

\smallskip\noindent\textbf{Fact~5. Bounding the number of possible attractor structures for sequential GDS maps.} The measure~$\kappa(G) = |\Acyc(G)/{\sim_\kappa}|$ is an upper bound for the number of distinct cycle structures that can be obtained by varying the update sequence~$\pi$. We can evaluate~$\kappa(G)$ in a manner similar to~$\alpha(G)$, that is,
\begin{equation}
 \kappa(G) = \kappa(G/e) + \kappa(G\setminus e) \;, \qquad e \in E(G) \text{ a cycle edge} \;.
\end{equation}
If~$G$ is cycle-free, then~$\alpha(G) = 1$. In particular, \emph{if~$G$ is a tree, there is only one cycle structure.}

\begin{example} Continuing the previous example sequence, we see that~$\kappa(G) = 4$. As before, take~$e = \{1,3\}$. Then~$\kappa(G/e) = 1$ and~$\kappa(G\setminus e) = 3$.
One cycle basis for~$G$ is~$\mathcal{C} = \bigl\{C_1 = (1,3,4,1), C_2 = (1,2,3,1) \bigr\}$. For example,~$\nu_\mathcal{C}(O(\pi=(1,2,3,4)))= (1,1)$ while~$\nu_\mathcal{C}(O(\pi' = (4,3,2,1))) = (-1,-1)$. We conclude that~$\pi$ and~$\pi'$ are not~$\kappa$-equivalent. However, we note that~$F_\pi$ and~$F_{\pi'}$ may still be cycle equivalent: that will depend on the particular choice of functions~$(f_i)_i$, a fact that motivates the next definition.
\end{example}
\begin{definition}
Let~$(f_i)_{i=1}^n$ be a sequence of vertex functions as above with dependency graph~$G$. Two acyclic orientations~$O(\pi), O(\pi') \in \Acyc(G)$ are \emph{cycle equivalent} if~$F_\pi$ and~$F_{\pi'}$ are cycle equivalent. Cycle equivalence classes are denoted~$[O(\pi)]_F$, and we let~$\kappa_F(G)$ denote the total number of cycle equivalence classes.
\end{definition}
Clearly, cycle-equivalence is a coarsening of $\kappa$-equivalence. By construction it follows that~$\kappa_F(G)\leq \kappa(G)$. The following is an open question: for a given (but arbitrary) graph~$G$, are there vertex functions~$(f_i)_i$ with dependency graph $G$ such that~$\kappa_F(G) = \kappa(G)$? 

Again $\alpha$- and $\kappa$-equivalence are strictly combinatorial constructions based on the structure of the dependency graph~$G$. However, they govern and restrict dynamics of corresponding GDS maps and significantly reduce the computational burden of many analysis problems. For example, to evaluate~$\kappa_F(G)$ we only need to determine the cycle structure across a set of $\kappa$-class representatives as in Fact~4. To get a sense of impact, consider the following example.
\begin{example}
Let $G$ be the binary hypercube of dimension~$3$ and consider some fixed but arbitrary sequence of vertex functions. In this case there are~$8!$ possible update sequences, we have~$\alpha(G) = 1862$ and~$\kappa(G) = 133$. To study cycle equivalence over~$G$, the above theory allows us to simplify the analysis by a factor of~$8!/133 > 300$. If the complete analysis over all~$8!$ update sequences would take~1 day, our approach would complete in less than~$5$ minutes.
\end{example}

\raggedbottom
\section{Overview of Method}
\label{sec:method} 

Algorithm~\ref{euclid} generates a complete set of~$\kappa$-class representatives for a graph~$G$. Note that a linear extension of an acyclic orientation~$O$ is a permutation~$\pi$ for which~$O_G(\pi) = O$. Using this set of representatives, the exploration of the possible cycle structures that can arise through varying the permutation update sequence now becomes more compelling. The algorithm uses Fact~4. Note that we choose a vertex~$v$ of maximal degree as that reduces the remaining computations. Again, we assume that~$G$ has been converted into the combinatorial graph~$G_c$ if~$G$ is not simple. Please note that the Python code and associated library that were used for the subsequent computations are available for download, see Section~\ref{sec:software} for details.

\begin{algorithm}
\caption{Complete set of $\kappa$-class representatives.}\label{euclid}
\begin{algorithmic}[1]
\State Select $v\in V(G)$ of maximal degree
\State $G' := G \setminus v$
\State $\textrm{kappaClassPermutationReps} := \emptyset$
\For {$O \in \textrm{AcyclicOrientations}(G')$} 
\State $ \textrm{flag = False}$
\For {$v' \textrm{ in } V(G')$}
\If{$\{v', v\} \not\in E(G) \And v' \in \textrm{SourceVertices}(O)$}
\State $\textrm{flag = True}$
\State $\textbf{break}$
\EndIf
\EndFor
\If{\textbf{not } flag }
\State kappaClassPermutationReps.Insert( Concatenate( $v$, LinearExtension($O$) ) )
\EndIf
\EndFor
\State $\textbf{return } \textrm{kappaClassPermutationReps}$
\end{algorithmic}
\end{algorithm}

We illustrate the algorithm using Example~\ref{ex:psEX} from Section~\ref{sec:back} for reference. Here the dependency graph~$G$ shown in Figure~\ref{fig:psExample} is simple. 
To construct all the acyclic orientations of~$G$ where $v$ is a unique source, that is, $\Acyc_v(G)$, form the graph $G'$ (delete $v$ from $G$) and start by constructing all its possible orientations (there will be $2^N$ of these, where $N$ is the number of edges of $G'$); discard the ones that are cyclic. For this, a graph library that can perform a topological sorting of directed graphs, e.g. NetworkX~\cite{hagberg-2008-exploring}, can be very useful. If an orientation is cyclic, the topological sort will fail and the orientation can be discarded.
However, if the orientation~$O'$ is acyclic and contains no source vertex~$v'$ such that~$\{v,v'\} \not\in E(G)$, we  form $O\in\Acyc_v(G)$ by adding~$v$ to~$O'$ and reintroducing all deleted edges incident to~$v$ orienting them so that~$v$ is a source. From a corresponding collection of update sequence representatives we can construct the cycle structures of the phase spaces $\Gamma(F_\pi)$. 

The algorithm returns a complete set of~$\kappa$-class representatives represented by update orders. If we compute~$F_{\pi}$ for all representative update orders~$\pi$ in this collection, we obtain all possible cycle structures of the SDS. In Example~\ref{ex:psEX}, we find~$4$ $\kappa$-class representatives. Examining the maps~$F_\pi$ for these~$4$ cases we see that the only possible cycle structure is one with two fixed points, namely~$(0,0,0,0)$ and~$(1,1,1,1)$ as depicted in Figure~\ref{fig:psExample}.   
        
\section{Biological System Models}
\label{sec:models}
In this section we present the biological network models that will be used in the computational analysis. In this biological context, the vertex functions, which are listed in Figures~\ref{fig:lacoperonfunctions} and~\ref{fig:CEnetworkfunctions}, represent regulatory interactions between molecules. The dependency graphs for these networks seen in Figures~\ref{fig:lacoperon} and~\ref{fig:CEnetwork} are directed and may contain loops. For simplicity, we have relabeled the vertices.

\subsection{Network Model For \emph{Lac} Operon}
\begin{figure}[htbp]
  \begin{center}
  \includegraphics[scale=.5]{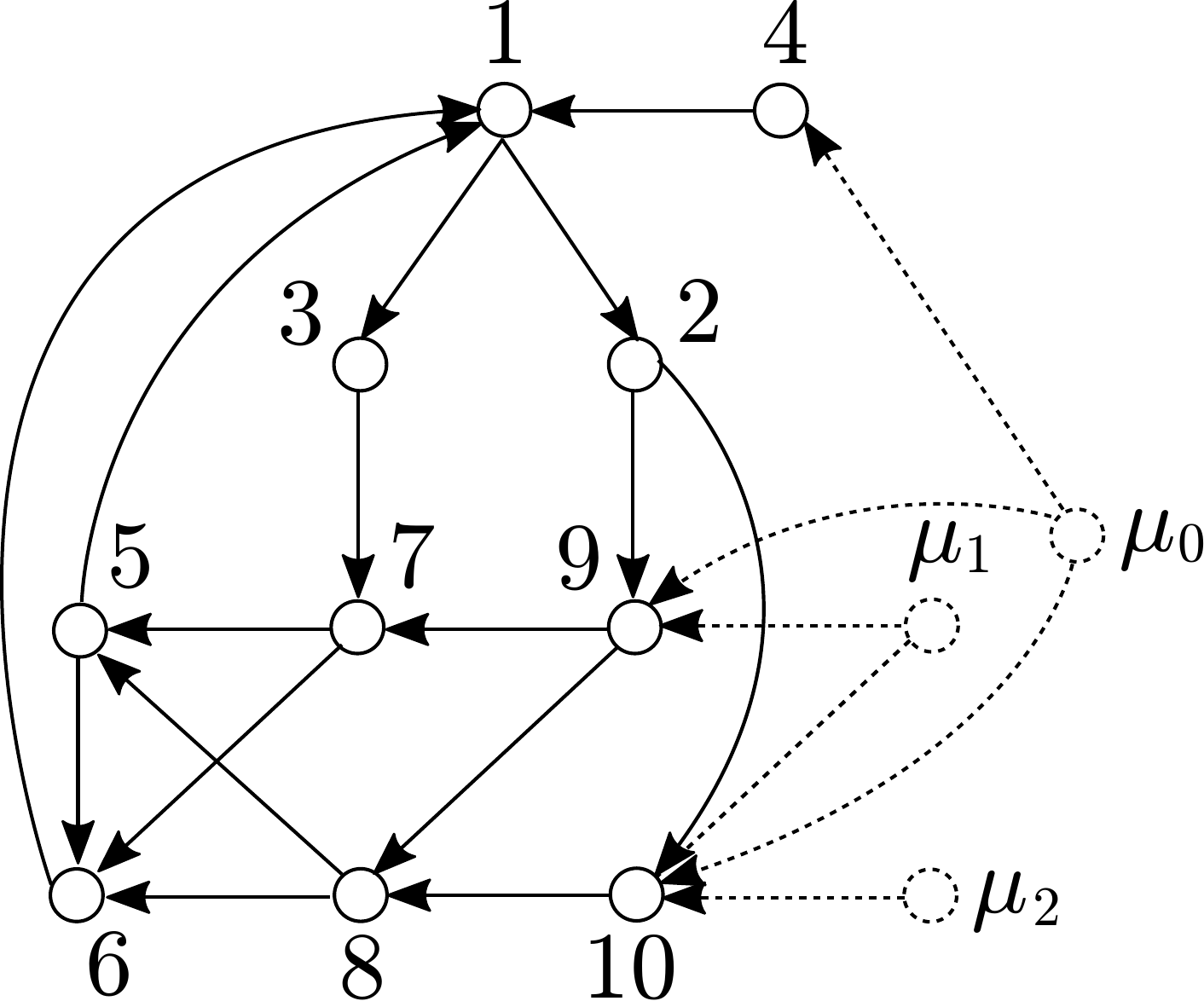}
  \caption{The dependency graph~$G$ for the network model in the~\emph{lac} operon (relabeled and reproduced from~\cite{veliz2011boolean}). Here $\mu_0$, $\mu_1$, and $\mu_2$ are parameters.}
  \label{fig:lacoperon}
  \end{center}
\end{figure}

The~\emph{lac} operon in~\emph{Escherichia coli} is the system responsible for the metabolism of lactose in the absence of glucose and is known to exhibit bistability, in the sense that the operon is either induced (ON) or uninduced (OFF)~\cite{montalva14attraction}. In~\cite{veliz2011boolean} the authors proposed a Boolean model for it, see Figure~\ref{fig:lacoperon}. 

\begin{figure}[H]
\begin{center}
\begin{tabular}{ccc} 
  
 {$\!\begin{aligned} 
               f_{1} &= x_4 \wedge \neg x_5 \wedge \neg x_6 \\    
               f_{2} &= x_1 \\
               f_{3} &= x_1 \\
               f_{4,\mu_0} &= \neg \mu_0 \\
               f_{5} &= \neg x_7 \wedge \neg x_8 \end{aligned}$} & $\, \, \, \, \,$  {$\!\begin{aligned} 
               f_{6} &= (\neg x_7 \wedge \neg x_8) \vee x_5 \\   
               f_{7} &= x_9 \wedge x_3 \\
               f_{8} &= x_9 \vee x_{10} \\
               f_{9,\mu_0,\mu_1} &= x_2 \wedge \mu_1 \wedge \neg \mu_0 \\
               f_{10,\mu_0,\mu_1,\mu_2} &	= \bigl((\mu_2 \wedge x_2 ) \vee \mu_1\bigr) \wedge \neg \mu_0 \\
               \end{aligned}$}

\end{tabular}
\end{center}
\caption{The vertex functions for the network model in the~\emph{lac} operon.}
\label{fig:lacoperonfunctions}
\end{figure} 


Here~$\mu_0 \in \{0,1\}$ and the pair~$(\mu_1,\mu_2)$ can be in three states; low, medium, or high which is represented by~$(\mu_1,\mu_2) = (0,0)$, $(0,1)$, and $(1,1)$ respectively. 

The authors in~\cite{montalva14attraction} consider dependency graph~$G$ under a parallel update and only find fixed points which represent the steady ON or OFF states of the operon. For more information on the biological interpretations of this network model see~\cite{veliz2011boolean}. We note that comprehensive analyses of this model under all sequential updates have been studied previously by Goles et al. in~\citep{montalva14attraction}.


\subsection{Network Model in \emph{Caenorhabditis Elegans}} 
Here we introduce the regulatory network model involved in the control of the cell cycle and cell differentiation in the~\emph{C. elegans} vulva precursor cells proposed in~\cite{weinstein2015model}. This model is, to our knowledge, the first model to include the molecular mechanism involved in the control of the postembryonic cell cycle of~\emph{C. elegans}.



\begin{figure}[htbp]
  \begin{center}
  \includegraphics[scale=.5]{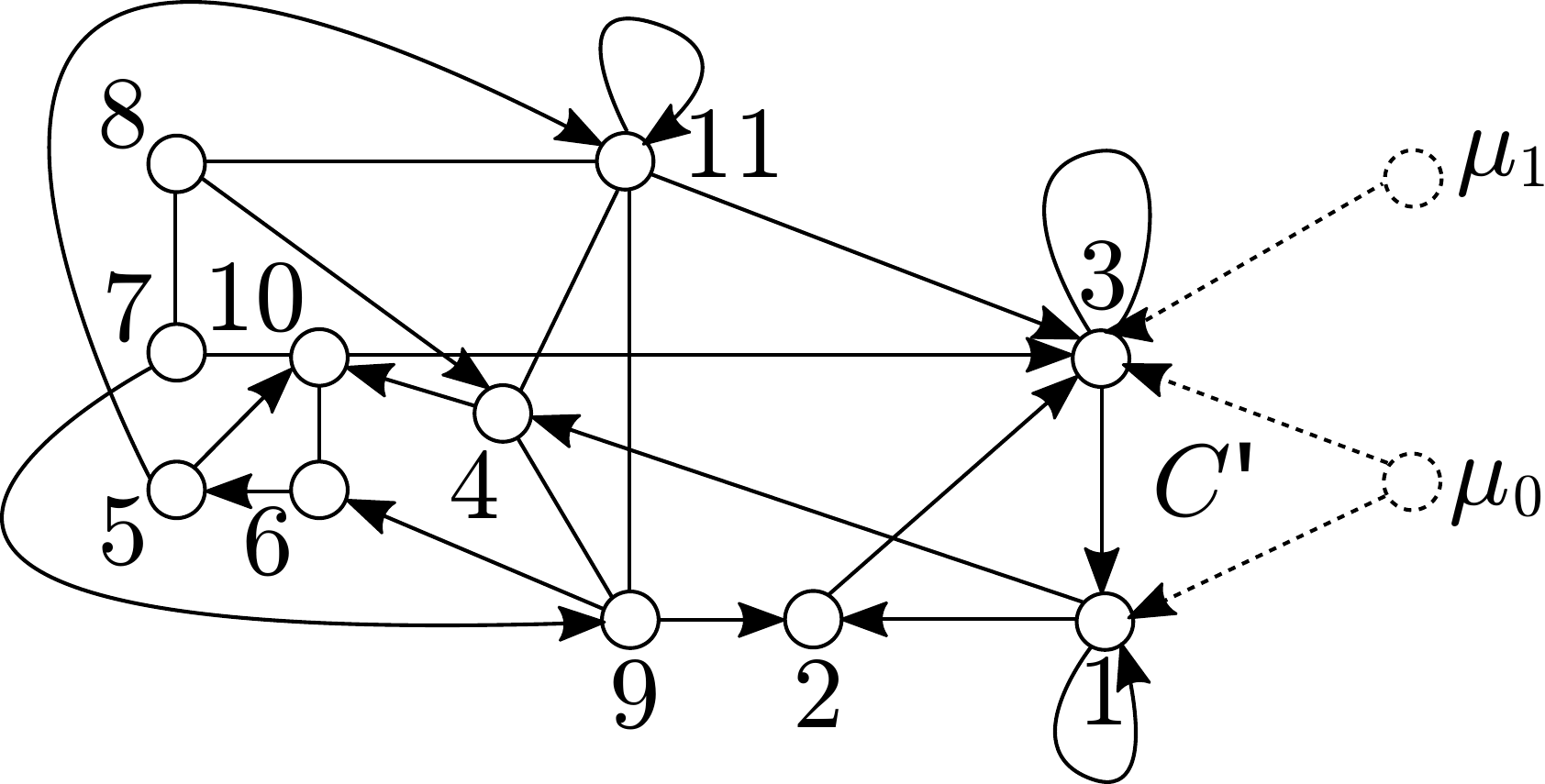}
  \caption{The dependency graph~$G$ for the network model in the~\emph{C. elegans} (relabeled and reproduced from~\cite{weinstein2015model}). We use~$G'$ to refer to the extended graph that includes the two dashed vertices and three dashed edges, which introduce the cycle $C'$.}
  \label{fig:CEnetwork}
  \end{center}
\end{figure}

In Figure~\ref{fig:CEnetwork} we show the dependency graph $G$. We denote the parameters by~$\mu_0,\mu_1$ where~$\mu_0 \in \{0,1,2,3\}$ and $\mu_1 \in \{0,1\}$.  

\begin{figure}[H]
\begin{center}
\begin{tabular}{ccc} 
  
 {$\!\begin{aligned} 
               f_{1,\mu_0} &=
				\begin{cases}
					2, &\text{if } (\mu_0 = 3 \wedge x_1 >0)\, \vee \\
					&(\mu_0 = 2 \wedge x_3 = 0 \wedge x_1 >0) \\
					0, &\text{if } (x_1<2) \, \wedge \\
					&\bigl((\mu_0 = 1 \wedge x_3 = 1) \vee \mu_0\bigr) \\
					1, &\text{otherwise}\\
				\end{cases} \\    
               f_{2} &= 
				\begin{cases}
					1, &\text{if } x_1 \leq 1 \vee x_9 = 1 \\
					0, &\text{otherwise}\\
				\end{cases} \\
               f_{3,\mu_0,\mu_1} &= 
                \begin{cases}
					1, &\text{if } \bigl((\mu_1=1 \wedge x_2=1)\, \vee\\
					&(x_3 \vee \mu_0)\bigr) \wedge (x_{10} \vee x_{11}) \\
					0, &\text{otherwise}\\
				\end{cases} \\
               f_{4} &= 
                \begin{cases}
					1, &\text{if } (x_1=0)\, \wedge \\
					&(x_9 = 0 \wedge x_8 =1 \vee x_{11}=1) \\
					0, &\text{otherwise}\\
				\end{cases} \\
               f_{5} &= 
                \begin{cases}
					1, &\text{if } x_6=0\\
					0, &\text{otherwise}\\
				\end{cases} \\

               \end{aligned}$} & $\, \,$  {$\!\begin{aligned} 
			   f_{6} &= 
                \begin{cases}
					1, &\text{if } x_9=0 \wedge x_{10}=0\\
					0, &\text{otherwise}\\
				\end{cases} \\

               f_{7} &= 
                \begin{cases}
					1, &\text{if } x_8=0 \wedge x_{10}=1\\
					0, &\text{otherwise}\\
				\end{cases} \\   
				
               f_{8} &= 
                \begin{cases}
					1, &\text{if } x_7=0 \wedge x_{11}=1\\
					0, &\text{otherwise}\\
				\end{cases} \\
			   
			   f_{9} &= 
                \begin{cases}
					1, &\text{if } x_4=0\, \wedge \\ 
					&x_7=0 \wedge x_{11}=0\\
					0, &\text{otherwise}\\
				\end{cases} \\
			   
			   f_{10} &= 
                \begin{cases}
					1, &\text{if } x_5=1 \wedge x_6=0\, \wedge \\
					&x_4=0 \wedge x_7=0\\
					0, &\text{otherwise}\\
				\end{cases} \\
			  
			   f_{11} &= 
                \begin{cases}
					1, &\text{if } x_4=1 \wedge x_8=0\, \wedge\\
					&x_5=1 \wedge (x_9=0 \vee x_{11}) \\
					0, &\text{otherwise}\\
				\end{cases} \\
			
               \end{aligned}$}

\end{tabular}
\end{center}
\caption{The list of vertex functions for the network model of~\emph{C. elegans}.}
\label{fig:CEnetworkfunctions}
\end{figure} 

The larger graph~$G'$ is the result of extending $G$ to include the parameters/vertices~$\mu_0,\mu_1$ and all their incident dashed edges. For the system over~$G'$ we write~$f_{3,\mu_0,\mu_1},f_{1,\mu_1}$ to $f_3,f_1$ and let the parameter vertex functions $f_{12}$ and $f_{13}$ simply  be the identify functions returning their own state. As a consequence of this, the phase space of $F$ can then be obtained as the disjoint union of the eight phase spaces of~$(F_{\mu_0,\mu_1})_{\mu_0,\mu_1}$ over~$G$. To assess dynamical diversity and equivalences we therefore have two graphs to consider,~$G$ and~$G'$. The advantage of using parameters is to cut down on memory requirements and allow for a compact discussion of results in Section~\ref{sec:results}.  

The authors in~\cite{weinstein2015model} consider the dependency graph~$G'$ under a parallel update and interpret periodic~$n$-cycles found as the patterns of molecular activation of the three vulval fates that cycle through the cell cycle. For more information on vulval fates or the cell differentiation process see~\cite{weinstein2015model}. 



\section{Results }
\label{sec:results}
The dependency graphs for the biological network models seen in
Figures~\ref{fig:lacoperon} and~\ref{fig:CEnetwork} are directed and contain
loops. However, as described in Section~\ref{sec:back}, we consider the combinatorial graph obtained by removing all loops and replacing each edge $(a,b)$ and pairs of edges $(a,b)$, $(b,a)$ by an undirected edge $\{a,b\}$.


\subsection{Network Model for \emph{lac} operon} 
The possible dynamics of this network model have already been studied comprehensively in~\cite{montalva14attraction}. In the following we will only consider the case where $\mu_0~=~\mu_1~=~0$~and~$\mu_2=1$, as it is shown in~\cite{montalva14attraction} that all other parameter and update schedule configurations reach a single fixed point. 

We extend and complement this work by utilizing the $\kappa$-class representative method to explicitly list all possible long-term dynamical structures of $F_{0,(0,1)}$ under sequential updates using the dependency graph $G$, see Table~\ref{tab:lacOpmultisets}. The two steady states (fixed points) found are $(0, 0, 0, 1, 1, 1, 0, 0, 0, 0)$ and $(1, 1, 1, 1, 0, 0, 0, 1, 0, 1)$, which correspond to the operon being OFF and ON, respectively.   

Using Algorithm 1, we find~$\alpha(G) = 14112$ and ~$\kappa(G) = 344$. We compute the phase space for each of these $344$ representative updates and find that~$\kappa_{F_{0,(0,1)}}(G) = 4$. The efficiency of pre-computing~$\kappa$-classes is clear here, as we can extract cycle structure results using only~$344$ update order representatives instead of $10!$ or even ~$14112$. The low ratio~$\kappa_{F_{0,(0,1)}}(G)/\kappa(G)$ provides an initial measurement on the update order robustness of the network. A low ratio would suggest a high degree of cycle structure preservation across different update permutations, a general characteristic of network robustness. 


\begin{table}[H]
\begin{center}
\begin{tabular}{ |c|c|c|c| } 
 \hline 
 \multicolumn{4}{|c|}{$\text{cycle structure multiset}: \text{frequency of $\kappa$-classes in cycle equivalence class}$} \\ [0.5ex]
 \hline
$\{1(2)\} : 263$ & $\{1(2), 2(1)\} : 31$ & $\{1(2), 3(2)\}: 31$ & $\{1(2), 2(1),4(3)\}: 19$ \\
 \hline 
\end{tabular}
\end{center}
\caption{The results displayed represent the cycle structure multisets for all $\kappa$-classes using the fixed parameters~$\mu_0 = \mu_1 = 0$ and $\mu_2=1$.}
\label{tab:lacOpmultisets}
\end{table}

\subsection{Network Model in \emph{Caenorhabditis Elegans}}
In the following we consider the two dependency graphs of this network model denoted by~$G$ and~$G'$, see Section~\ref{sec:models}. Here we show how to relate~$\alpha$- and~$\kappa$-classes over these two graphs.

For~$\alpha$-equivalence we note that each acyclic orientation $O(\pi)\!\!\in\!\!\Acyc(G)$ extends to precisely six acyclic orientations~$O_1(\pi_1),\dots, O_6(\pi_6) \in \Acyc(G')$ where all maps~$O_i(\pi_i)$ are equivalent when restricted to~$E(G)$. Consequently, $\alpha(G') = 6\alpha(G)$. However, all linear extensions~$\pi_i$ from these six orientations result in equivalent dynamics under~$F_{\pi_i}$. 

For $\kappa$-equivalence we use Coleman's $\nu$-function (see Section~\ref{sec:back}). Let $\mathcal{C} = (C_1,\dots,C_m)$ be a cycle basis for $G$. We extend $\mathcal{C}$ to a cycle basis for $G'$ by adding $C'$ to get $\mathcal{C'} = (C_1,\dots,C_m,C')$, see Figure \ref{fig:CEnetwork}. Since $\nu_{C'}(O')\!\!\in\!\!\{-1,1\}$ for all~$O'\!\!\in\!\!\Acyc(G')$, it follows that each~$\kappa$-class over $G$ extends to precisely two $\kappa$-classes over $G'$.  Consequently, $\kappa(G') = 2\kappa(G)$. The six orientations of $G$ seen in the paragraph above are distributed across two~$\kappa$-classes over $G'$ according to their value under $\nu_{C'}$. However, due to the functional equivalence of the six orientations, these two distinct~$\kappa$-classes must be contained in the same cycle equivalence class over $G'$. 

We conclude that each~$\kappa$-class over~$G$ gives rise to two distinct and cycle equivalent~$\kappa$-classes over~$G'$ and each~$\alpha$-class over~$G$ gives rise to six distinct and functionally equivalent~$\alpha$-classes over $G'$. To find all possible cycle structures using~$G'$ it is sufficient to evaluate each~$\kappa$-class representative using~$G$ under all eight parameter configurations, see Section~\ref{sec:models}. 

We first consider dynamics over the dependency graph~$G$ and find~$\alpha(G) = 158208$ and~$\kappa(G) = 5312$ using Algorithm 1. These are graph combinatorial measures which are independent of the parameter choice and functions, however altering the parameters $\mu_0$ and $\mu_1$ may change the dynamics of the system and therefore change the measure~$\kappa_{F_{\mu_0,\mu_1}}\!\!(G)$, see Table~\ref{tab:kappaFG}. 



\begin{observation}
Let $\pi \in S_G$ and parameters $(\mu_0,\mu_1)\in{\{0,1,2,3\}\times\{0,1\}}$ be given and fixed. The resulting phase space~$\Upgamma(F_{(\mu_0,\mu_1),\pi})$ is restricted to the following long-term dynamics:
\begin{enumerate}
\item A single~$n$-cycle, or
\item Two distinct~$n$-cycles with the same length~$n$ (bistability). 
\end{enumerate}
In both cases we have~$3 \leq n \leq 10$. If $(\mu_0,\mu_1)\in{\{(0,1),(1,0),(1,1),(3,1)\}}$ or a parallel update is used, we only obtain a single~$n$-cycle (case 1). 
\label{obs:bistability}
\end{observation}

In Table~\ref{tab:kappaFG} we fix the parameters~$(\mu_0,\mu_1)$ and for each combination, we display~$\kappa_{F_{\mu_0,\mu_1}}\!\!(G)$ and the frequency of $\kappa$-classes that resulted in bistable cycle structures (two distinct~$n$-cycles with the same length~$n$). We note that with the use of a parallel update, we obtain a single~$n$-cycle for any parameter combination (bistability is not present). 

\begin{table}[H]
\begin{center}
\begin{tabular}{ |c|c|c| } 
 \hline 
 $(\mu_0,\mu_1)$ & $\kappa_{F_{\mu_0,\mu_1}}\!\!(G)$  & $\text{frequency of $\kappa$-classes that result in bistability}$  \\ 
 \hline
$(0,0)$ & $14$ & $1695$\\
$(0,1)$ & $8$ & $0$\\
$(1,0)$ & $8$ & $0$\\
$(1,1)$ & $8$ & $0$\\
$(2,0)$ & $12$ & $1664$\\
$(2,1)$ & $8$ & $84$\\
$(3,0)$ & $12$ & $1664$\\
$(3,1)$ & $7$ & $0$\\
 \hline

\end{tabular}
\end{center}
\caption{For each fixed parameter combination~$(\mu_0,\mu_1)$ we display $\kappa_{F_{\mu_0,\mu_1}}\!\!(G)$ and the frequency of $\kappa$-classes that result in bistable cycle structures. In each case there are~$5312$ total~$\kappa$-classes.}
\label{tab:kappaFG}
\end{table}


For the extended dependency graph~$G'$ we find $\alpha(G') = 6\alpha(G) = 949248$, $\kappa(G') = 2\kappa(G')= 10624$, and~$\kappa_F(G') = 125$. In Table~\ref{tab:multisets} we showcase some of the variety of cycle structures from~$G$ by listing several cycle structure multisets found. The multisets found in Table~\ref{tab:multisets} correspond to the cycle structures of some $4394$ $\kappa$-classes. The frequency of a cycle structure multiset is the number of $\kappa$-classes that resulted in the same cycle structure displayed by the multiset. We can then group the $\kappa$-classes that resulted in the same cycle structure to form a cycle equivalence class. Statistics of multisets over all~$10624$ $\kappa$-classes are found in Table~\ref{tab:statistics}.

\begin{table}[H]
\begin{center}
\begin{tabular}{ |l|l|l| } 
 \hline 
 \multicolumn{3}{|c|}{$\text{cycle structure multiset}: \text{frequency of $\kappa$-classes in cycle equivalence class}$} \\ [0.5ex]
 \hline
 $\{3(11)\} : 1094$ & $\{4(4), 5(4)\} : 914$ & $\{4(4), 5(2), 6(2)\}: 230$ \\
 $\{4(11)\}: 524$ & $\{5(4), 6(4)\}: 662$ & $\{5(4), 6(2), 7(2)\}: 200$ \\  
 $\{7(9)\}: 118$& $\{4(5), 5(4)\} : 204$ & $\{7(4), 8(2), 9(2)\}: 70$ \\
 $\{6(9)\}: 132$ & $\{3(6),7(5)\}: 196$ & $\{7(4), 9(2), 10(2)\}: 50$ \\   
\hline
\end{tabular}
\end{center}
\caption{The results displayed represent the cycle structure multisets for~$4394$ of the~$10624$ total $\kappa$-classes over~$G'$.  }
\label{tab:multisets}
\end{table}

\begin{figure}[htbp]
  \begin{center}
  \includegraphics[scale=.3]{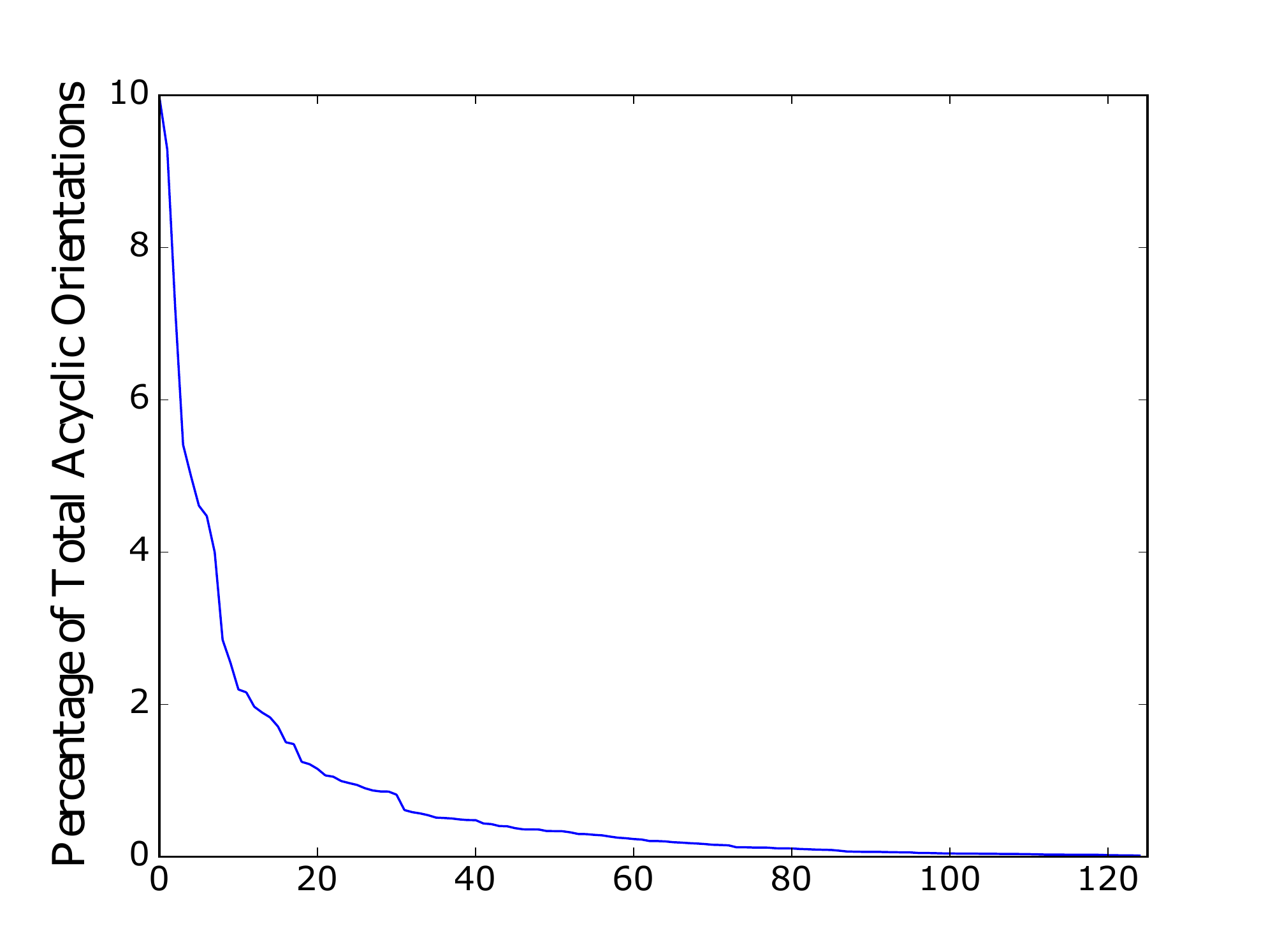}
  \caption{Distribution of all acyclic orientations of $G'$ across all cycle equivalence classes. On the horizontal axis we enumerate all~$125$ cycle equivalence classes and plot, in decreasing order, the percentage of total acyclic orientations contained in a corresponding class.}
  \label{fig:AcycDist}
  \end{center}
\end{figure}

To gain further initial insight on the update robustness of the model, we may visualize how acyclic orientations are distributed across cycle equivalence classes. In Figure~\ref{fig:AcycDist} we enumerate all cycle equivalence classes and plot, in decreasing order, the percentage of total acyclic orientations contained in a corresponding class. In a robust network model we expect this graph to be heavily front-loaded, as we expect a large percentage of possible dynamics to result in the same cycle structures.

\raggedbottom

We find that around $75\%$ of all acyclic orientations are distributed across~$23$ cycle equivalence classes. To further discuss results in the context of system robustness, let us consider the cycle structure results found in~\cite{weinstein2015model} when the update is a parallel one. The cycle structure multiset for the parallel update is~$\{9(4),10(2),11(2)\}$. If we assume the model has robustness across update order variations, then we expect cycle structure multiset data to remain similar in the context of the biological network. Finding and interpreting these similarities can require a vast amount of insight on the biological system. 

\begin{table}[H]
\begin{center}
 \begin{tabular}{|c|c|c|c|} 
 \hline 
 \multicolumn{4}{|c|}{$\text{total number of cycles (size of multiset)$\colon$ frequency of $\kappa$-classes}$} \\ [0.5ex]
 \hline
 $8 \colon 6496$ & $9 \colon 800$ & $10 \colon 570$ & $11 \colon 2758$ \\ 
 \hline 
\end{tabular}

\end{center}
\caption{The frequency of~$\kappa$-classes that result in cycle structures with the same total number of~$n$-cycles (same multiset size).}
\label{tab:statistics}
\end{table}

In Table~\ref{tab:statistics} we see that the total number of cycles within a cycle structure multiset is roughly preserved from the parallel update scheme, as~$61.1\%$ of all $\kappa$-classes result in multisets of size eight. We also find that within a multiset there are at most three distinct $n$-cycle lengths where the length~$n$ is at least three and at most ten.  

\section{Software}
\label{sec:software}

The software used and referenced in this paper is available from GitHub (\url{https://github.com/HenningMortveit/gds-framework-python}) under the Artistic License 2.0 (\url{https://opensource.org/licenses/Artistic-2.0}). The source has been developed and tested under Python version~2.7. All computational examples in the text can be found in \texttt{./biographs-paper.py}. By inspecting the system models in \texttt{./gds/biographs.py}, one can adapt these to other systems and analyses.

\section{Conclusions and Future Work}
\label{sec:conc}
In this paper we demonstrated a computationally efficient method for finding all possible cycle structures arising in asynchronous biological models  as in Equation~\ref{eq:sds}. In summary, we utilized $\kappa$-equivalences over the dependency graph~$G$ to precompute classes of permutations whose corresponding SDS maps are cycle equivalent. Therefore, computation over one representative update schedule per~$\kappa$-class is sufficient for comprehensive analysis of possible cycle structures. This analysis provides immediate feedback to biological modelers about the presence and diversity of possible long-term dynamical structures resulting from varying the update order and the robustness of their model. For example, in the network model of~\emph{C. elegans} we saw that the model is bistable under select sequential updates and fixed parameters. Additionally, in the~\emph{lac} operon model we find that certain $n$-cycles are admissible. In the context of this biological model, these limit cycles have no clear interpretation. However, from the frequency distribution of $\kappa$-classes in Table ~\ref{tab:lacOpmultisets}, we see that these cycles are relatively uncommon. 

Aside from efficiency, this method also provides a more refined approach to answering the general question of determining how the dynamics of a model change over different updating schemes. In this paper, we have simplified the question to determining the variety and distribution of cycle structures across different sequential updates. We are able to compare the total number of distinct cycle structures of a model, that is~$\kappa_F(G)$, based on a worse-case scenario arising from the bound~$\kappa(G)$. In the case of overall attractor stability, we expect this ratio~$\kappa_F(G)/\kappa(G)$ to be small and expect a small variety of possible cycle structures. We see that this is clearly the case for the network model in the~\emph{lac} operon, as~$10! = 3628800$ permutation update order dynamical systems (using~$\mu_0 = \mu_1 = 0$ and $\mu_2=1$) will result in only four possible cycle structures. We see a similar case for the network model in the~\emph{C. elegans}, as~$13! = 6227020800$ permutation update orders will result in~$125$ possible cycle structures. 

One future direction of work would be to analyze the possible periodic states within the long-term dynamical structures. The results of such analysis may be better interpreted in the context of the biological system model and could provide additional feedback to modelers. For example, we may use this feedback to further investigate and interpret the cases where the network model in~\emph{C. elegans} is bistable. We claim that comprehensive information of this type is also efficiently achievable in an approach that utilizes~$\kappa$-equivalence. 

In this work, we have restricted ourselves to asynchronous updates using permutations sequences. In ongoing work, we are extending this theory to  $\kappa$-equivalence for \emph{block-sequential} update orders. While this adds complexity, exploring stability of biological models in this more general setting may also allow additional insight on sensitivity and robustness. 

Finally, we mention that frequency of $\kappa$-classes, or more precisely, the frequency of cycle structures (as multisets) observed across $\kappa$-classes, gives initial insight into the respective likelihoods of attractors. A more elaborate and computationally demanding analysis may consider the following: each $\kappa$-class is a set of acyclic orientations, and each acyclic orientation has associated a set of linear extensions, that is, the update sequences that map to that particular orientation using Equation~\eqref{eq:phi_map}. For a particular biological system, there may be a statistical distribution across the set of permutation update sequences, and one may choose to use this when inducing weights of the $\kappa$-classes. We have chosen to leave this analysis for future work. 

\section*{Acknowledgments}

We thank our external collaborators and members of the Network Dynamics and Simulation Science Laboratory (NDSSL) for their suggestions and comments.  This work has been partially supported by DTRA Grant HDTRA1-11-1-0016.

\bibliographystyle{spmpsci}      
\bibliography{bibliofile}   
\end{document}